\documentclass[aps,prl,twocolumn,superscriptaddress,floatfix]{revtex4}
\usepackage{graphicx,color}
\usepackage{dcolumn}
\usepackage{hyperref}
\usepackage{amsmath}
\hfuzz=\maxdimen
\usepackage{float}
\begin{document}

\title{Investigation of the Pauli paramagnetic effect in systematically tuned NbN thin films}


\author{Xiaoni Wang}
\affiliation{Laboratory of Superconducting
Electronics, Shanghai Institute of Microsystem and Information
Technology, Chinese Academy of Sciences, Shanghai 200050,
China}\affiliation{University
of Chinese Academy of Science, Beijing 100049, China}

\author{Lijie Wang}
\affiliation{Department of Physics and State Key Laboratory of Surface Physics, Fudan University, Shanghai 200433, China}

\author{Yixin Liu}
\affiliation{Laboratory of Superconducting
Electronics, Shanghai Institute of Microsystem and Information
Technology, Chinese Academy of Sciences, Shanghai 200050,
China}\affiliation{University
of Chinese Academy of Science, Beijing 100049, China}

\author{Wanpeng Gao}
\affiliation{Laboratory of Superconducting
Electronics, Shanghai Institute of Microsystem and Information
Technology, Chinese Academy of Sciences, Shanghai 200050,
China}\affiliation{University
of Chinese Academy of Science, Beijing 100049, China}

\author{Yu Wu}
\affiliation{Laboratory of Superconducting
Electronics, Shanghai Institute of Microsystem and Information
Technology, Chinese Academy of Sciences, Shanghai 200050,
China}

\author{Zulei Xu}
\affiliation{Laboratory of Superconducting
Electronics, Shanghai Institute of Microsystem and Information
Technology, Chinese Academy of Sciences, Shanghai 200050,
China}\affiliation{University
of Chinese Academy of Science, Beijing 100049, China}

\author{Hua Jin}
\affiliation{Laboratory of Superconducting
Electronics, Shanghai Institute of Microsystem and Information
Technology, Chinese Academy of Sciences, Shanghai 200050,
China}

\author{Lu Zhang}
\affiliation{Laboratory of Superconducting
Electronics, Shanghai Institute of Microsystem and Information
Technology, Chinese Academy of Sciences, Shanghai 200050,
China}

\author{Wei Peng}
\affiliation{Laboratory of Superconducting
Electronics, Shanghai Institute of Microsystem and Information
Technology, Chinese Academy of Sciences, Shanghai 200050,
China}\affiliation{University
of Chinese Academy of Science, Beijing 100049, China}

\author{Zhen Wang}
\affiliation{Laboratory of Superconducting
Electronics, Shanghai Institute of Microsystem and Information
Technology, Chinese Academy of Sciences, Shanghai 200050,
China}\affiliation{University
of Chinese Academy of Science, Beijing 100049, China}

\author{Wei Li}\email[]{w$_$li@fudan.edu.cn}
\affiliation{Department of Physics and State Key Laboratory of Surface Physics, Fudan University, Shanghai 200433, China}

\author{Gang Mu}
\email[]{mugang@mail.sim.ac.cn} \affiliation{Laboratory of Superconducting
Electronics, Shanghai Institute of
Microsystem and Information Technology, Chinese Academy of Sciences,
Shanghai 200050, China}\affiliation{University
of Chinese Academy of Science, Beijing 100049, China}

\author{Zhirong Lin}
\email[]{zrlin@mail.sim.ac.cn} \affiliation{Laboratory
of Superconducting
Electronics, Shanghai Institute of
Microsystem and Information Technology, Chinese Academy of Sciences,
Shanghai 200050, China}\affiliation{University
of Chinese Academy of Science, Beijing 100049, China}





\begin{abstract}
Superconductivity and the normal-state properties of NbN films can be tuned in a wide range, supplying a suitable platform to investigate the systematical evolution of the superconducting performances.
Herein, we report the upper critical field of NbN films in both the vertical ($B\perp$ film) and parallel ($B\parallel$ film) orientations over a wide temperature range.
Eight samples with the superconducting critical temperature $T_c$ ranging from 2.5 K to 9.8 K are studied. Meanwhile, the normal-state resistivity is tuned by more than six times by changing the conditions of the film growth.
It is found that the magnitudes of the upper critical field in both field directions ($B_{c2}^{\perp}$ and $B_{c2}^{\parallel}$) exceed the paramagnetic limiting field $B_p$.
The temperature dependent $B_{c2}^{\perp}$ can be described by the extended Werthamer--Helfand--Hohenberg (WHH) model considering the Pauli spin paramagnetism. Meanwhile, the $B_{c2}^{\parallel}$-$T$ data shows the feature of
two-dimensional superconductivity in the temperature range near $T_c$. The evolution of the obtained
Maki parameter with other parameters, such as the slope of the upper critical field near $T_c$ and the normal-state resistivity, are discussed to reveal the characteristics of the Pauli paramagnetic effect in
this system.


\end{abstract}


\maketitle


\section*{1 Introduction}
Niobium nitride (NbN) reveals a relatively high superconducting (SC) transition $T_c$ of about 17 K~\cite{Keskar1971,Oya1974}. This relatively high value of $T_c$ is a consequence of the strong electron-phonon interaction and
low density of states at the Fermi level~\cite{Oya1974}. With the continuous progress of the technique for the film growth~\cite{Wang1996},
recent years, thin films of NbN have been used in superconducting nanowire single photon detector~\cite{YouLX2017} and superconducting quantum interference device\cite{SQUID2017}.
Moreover, this material is also adopted in the investigations of quantum phase slips in ultrathin nanowires~\cite{QPS2022}.
The SC properties of this system can be affected by the slight deviation of its composition from the stoichiometric value~\cite{Oya1974,Wang1996}.
This provides a good opportunity to systematically investigate the normal-state and SC behaviors in a wide parameter range.

In the early studies~\cite{Chockalingam2008,Joshi2018}, it has been noticed that NbN film displays a rather high upper critical field.
However, an in-depth investigation on the mechanism behind is lacking. Moveover, the information with the
field parallel to the film surface and the anisotropic effect has not been studied.
In a type-II superconductor, the upper critical field is affected by the interactions between the external magnetic field and orbital motion of the SC electrons, effect of the
field on the electron spin magnetic moments (the so-called Pauli paramagnetic effect), and the spin-orbit scattering~\cite{Maki1964,Maki1966,WHH-1,WHH-2}.
A quantitative evaluation of these different origins of pair-breaking effect is important to understanding the intrinsic SC behavior of the NbN system.

In this study, we conducted an in-depth investigation on the upper critical field of NbN thin films under the magnetic field up to 12 T. In order to obtain the information of upper critical
field in a wide temperature range, the values of $T_c$ are tuned to the range below 10 K. The paramagnetic limiting field $B_p$ is exceeded by the upper critical field in both directions ($B\perp$ film and $B\parallel$ film).
The extended Werthamer--Helfand--Hohenberg (WHH) model considering both the orbital and Pauli spin paramagnetic effects can give a fine description to the out-of-film upper critical field $B_{c2}^{\perp}$.
The influence of spin-orbit impurity scattering
is unconspicuous. The Pauli paramagnetic effect is further discussed by examining the evolution of the Maki parameter in the systematically tuned samples.

\begin{figure*}
\includegraphics[width=18cm]{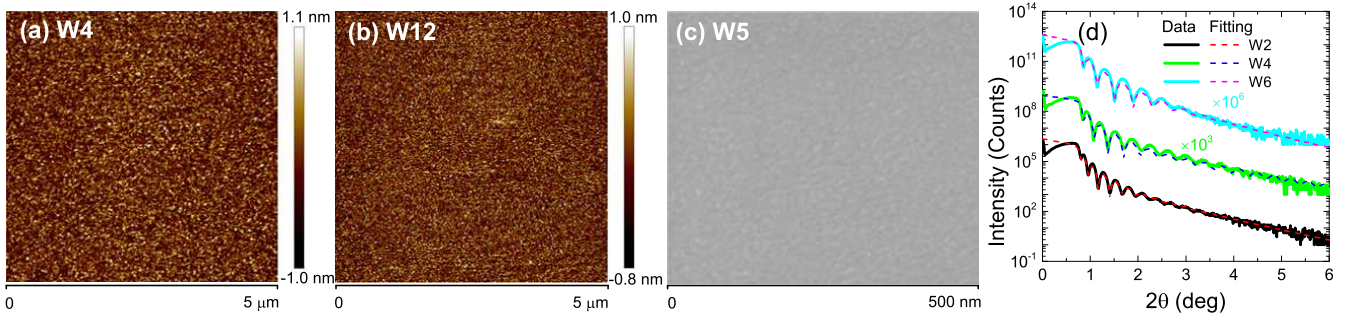}
\centering \caption {(a, b) AFM images of the samples W4 and W12, respectively. (c) SEM picture of sample W5. (d) X-ray reflection curve for three
typical samples. The dashed lines show the fitting results.}
\label{fig1}
\end{figure*}

\section*{2 Materials and methods}\label{sec:4}
The NbN thin films were deposited using an DC reactive magnetron sputtering equipped with a high vacuum pump, which supplies a high vacuum environment with a minimum base pressure of 5$\times10^{-8}$ Pa.
The deposition chamber consists of magnetron and Nb target powered by stabilized dc source. The high-resistivity Si was used as the substrate.
The ultrahigh vacuum in the chamber and the high purity of Nb target (99.99\%) in our experiment will reduce the contaminants and guarantee the high quality of the resultant films.
The sputtering process was carried out at room temperature in an mixed atmosphere of Ar (99.999\%) and N$_2$ (99.999\%). The magnetron sputtering we used has a small adjustable power range,
resulting in a rather slow rate of the film growth.
This is beneficial to a better control over the thickness and uniformity of the films. Moreover, the sample holder is constantly rotated to make the film more uniform during the process of film sputtering.

The N$_2$ content in the Ar/N$_2$ mixture and chamber pressure were changed systematically to tune the physical properties of the films. In addition, by regulating the film deposition time, the thickness of the films were controlled
within the range of 20-30 nm. The detailed parameters of the film growth are listed in Table 1.
The films were etched into the line shape with the dimension 10$\times$500 $\mu$m$^2$ using reactive ion etching (RIE) for the electrical transport measurements.

\begin{figure}
\includegraphics[width=7cm]{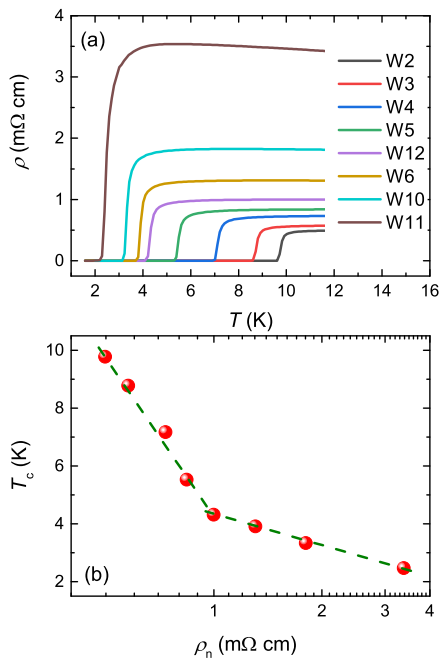}
\caption {(a) Temperature dependence of resistivity $\rho_n$ under zero field for eight NbN films. (b) Values of $T_c$ as a function of normal-state resistivity in the semi-logarithm scale.} \label{fig2}
\end{figure}

The surface morphologies of films were
measured by atomic force microscope (AFM, Bruker Dimension Icon) and scanning electron microscopy (SEM, Zeiss Gemini300).
The thicknesses of the films were determined by the X-ray reflectivity (XRR) measurements (High Resolution XRD, Bruker, D8 Discover).
The electrical transport measurements of samples W2, W3, W4, W5, W6, and W11 were
performed using a cryostat (Oxford Instruments TeslatronPT cryostat
system) with the field of up to 12 T. Other samples (W10 and W12) were measured on a physical property measurement system (PPMS, Quantum Design) with the magnetic field of up to
9 T. The applied
electric current is 0.5 $\mu$A during the transport measurements.
The magnetic field were applied at two different orientations
($B\perp$ film and $B\parallel$ film).
Current was applied perpendicular to the direction of the magnetic field.

\begin{figure*}
\includegraphics[width=13cm]{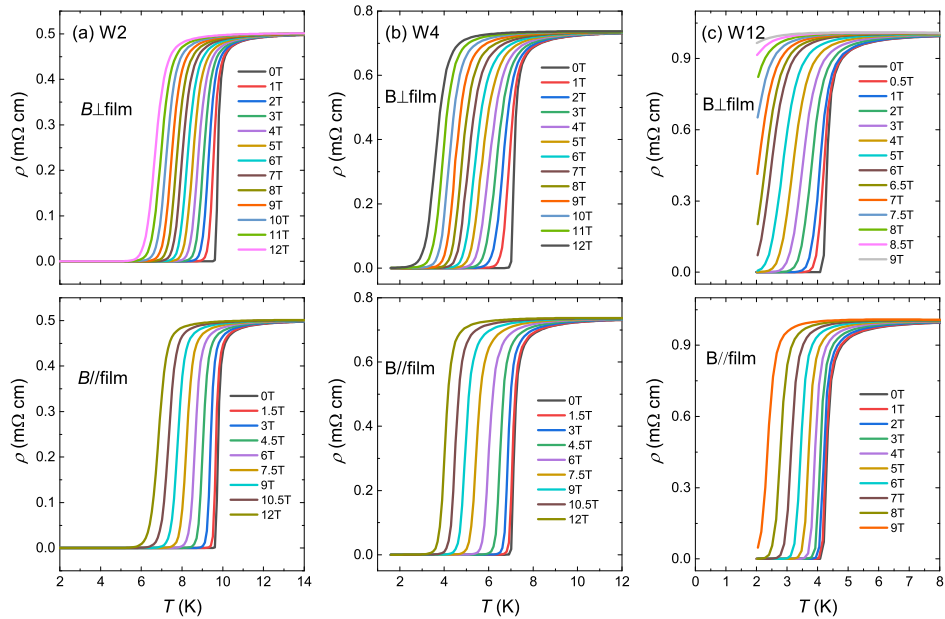}
\caption {Temperature dependent resistivity in different magnetic fields for the three films W2 (a), W4 (b), and W12 (c). The field is applied along two
different directions ($B \perp$ film and $B\parallel$
film).} \label{fig3}
\end{figure*}

\begin{table}
\centering
\caption{Parameters of the film growth and characterization for the
eight samples.}
\begin{tabular}
{p{1.2cm}<{\centering}p{1.8cm}<{\centering}p{1.2cm}<{\centering}p{1.5cm}<{\centering}p{1.2cm}<{\centering}p{1.2cm}<{\centering}}\hline \hline
  &   N$_2$ content   &  Pressure & Time & $d$  & $R_q$   \\
  Name &   (\%)   & (Pa) & (min)  & (nm) & (nm) \\
\hline
W2    & 21      & 0.11  &  20  & 29.5 & 0.31  \\
W3   & 25      & 0.11  &  20  & 26.7 & 0.22    \\
W4   & 32      & 0.12  &  20  & 22.9 & 0.30    \\
W5   & 41      & 0.12  &  20  & 21.1 & 0.29    \\
W12  & 50      & 0.10  &  20  & 22.0 & 0.25   \\
W6    & 50      & 0.13  &  20  & 20.1 & 0.24 \\
W10    & 50      & 0.16  &  30  & 27.0 & 0.14 \\
W11    & 50      & 0.20  &  30  & 25.0 & 0.25 \\
\hline
\hline
\end{tabular}
\label{tab.1}
\end{table}

\section*{3 Results}

\subsection*{3.1 Characterization}
Eight samples grown under different conditions are investigated in this work.
The surface morphologies of the films were measured by AFM. All the films show a root-mean-square roughness($R_q$) of less than 0.3 nm at the scan size of 5 $\mu$m, indicating a high level of flatness in our samples.
As an example, we show the AFM images of two samples W4 and W12 in Figs. 1(a) and (b), respectively.
It can be seen that the size of the film grain is about 20 nm with a uniform distribution. In addition, we used a scanning electron microscope to observe the surface topography of the film at a magnification of 160k, see Fig. 1(c).
The AFM and SEM images of other samples can be seen Figs. S1, S2 and S3 in the Supplementary Material (SM), both of which reveal the smooth and dense surface of the samples.

To get information about the thickness ($d$) of the films, we carried out the X-ray reflectivity measurements. The reflection curve was analyzed using LEPTOS software from Bruker.
By constructing a simple three-layer structure model of silicon substrate, NbN film, and the niobium oxide layer, the theoretical curve is obtained to fit with the measured reflection curve,
which is presented by the dashed lines in Fig. 1(d). The XRR data of other samples are shown in Fig. S4.
The obtained values of film thickness are listed in Table 1.

\subsection*{3.2 Superconducting transition}
Physical properties of the films in both the superconducting and
normal states are studied by the electrical transport measurements.
Fig. 2(a) shows the temperature dependent resistivity of
the eight films. Sharp SC transitions occurred in the low temperature region. The SC critical temperature $T_c$ reveals a systematical decrease with the increase of N$_2$ content and/or the chamber pressure.
Meanwhile, the magnitudes of resistivity in the normal-state also show a systematical evolution. In order to unveil the correlation between the SC and normal-state properties, we plot $T_c$ as a function
of $\rho_n$ in Fig. 2(b). Here $T_c$ is determined using the criteria of 50\%$\rho_n$ to eliminate the effect of SC fluctuation near the
onset of the SC transition. $\rho_n$ is the resistivity in the normal state at 14 K, which is slightly above $T_c$. It is clear that $T_c$ shows a linear dependence with logarithm of $\rho_n$ with a change in the
slope at round $\rho_n$ = 1 m$\Omega$ cm.

The behaviors of electrical transport under external magnetic field can supply very important information about field induced pair-breaking effect.
In Fig. 3, we show temperature-dependent resistivity with various
magnetic fields perpendicular and parallel to the film surface for the three samples. The data of other five samples can be seen in Figs. S5 and S6 in SM. As the magnetic field increases, the SC
transition shifts to lower temperatures for all the three samples. The field induced broadening effect is more significant with the vertical field ($B\perp$ film), indicating a more active flux motion in this field orientation.
In addition, the degree of suppression of magnetic field on the SC transition shows a certain degree of anisotropy, which will be discussed in the next section.

\begin{figure}
\includegraphics[width=7cm]{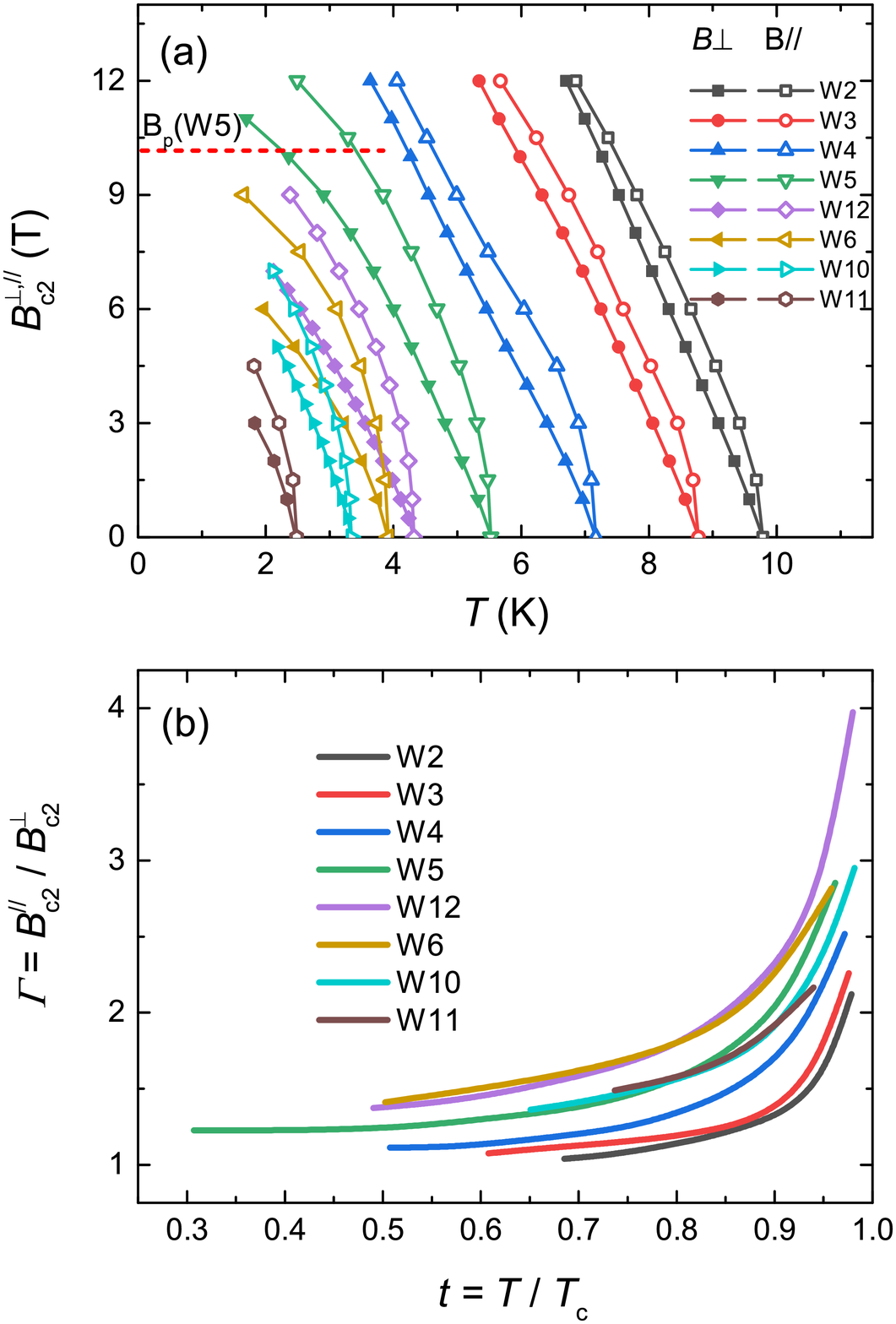}
\caption {(a) Upper critical field $B^{\perp,\parallel}_{c2}(T)$ in two field directions versus temperature for the eight NbN samples. The red dashed line index the position of the paramagnetic limiting field $B_p$ for sample W5.
(b) Anisotropic parameter $\Gamma=B{c2}^{\parallel}/B_{c2}^{\perp}$ as a function of the reduced temperature $t=T/T_c$.} \label{fig4}
\end{figure}

\begin{figure}
\includegraphics[width=9cm]{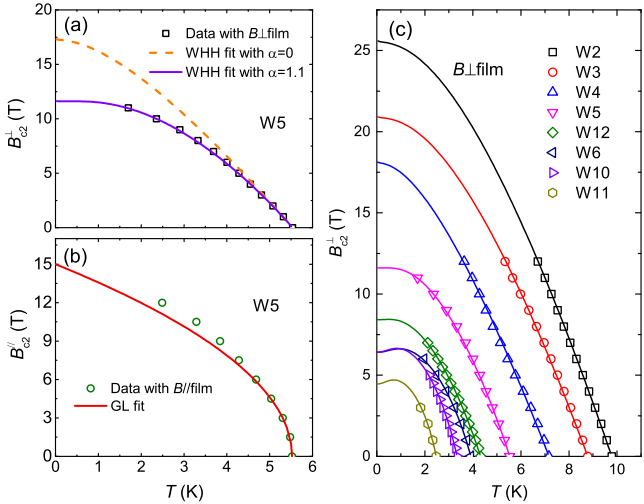}
\caption {(a) Temperature dependence of $B^{\perp}_{c2}(T)$ for sample W5. The dashed and solid lines are the extended WHH fits for the experimental data with different Maki parameters.
(b) Temperature dependence of $B^{\parallel}_{c2}(T)$ for sample W5. The red solid line is result of the 2D GL fit. (c) Temperature dependence of $B^{\perp}_{c2}(T)$ for all the eight samples. The solid lines reveal
the fitting results based on the extended WHH theory.} \label{fig5}
\end{figure}

\subsection*{3.3 Upper critical field}
The upper critical fields in the two orientations, $B_{c2}^{\perp}$ and $B_{c2}^{\parallel}$, for all the eight samples as a function of temperature are shown in Fig. 4(a).
As we have mentioned, the criteria of 50\%$\rho_n$ is adopted in determining the values of upper critical field. All the samples show the rather large values of the upper critical field, which
exceeds the paramagnetic limiting field $B_p$ (= 1.84 $\times T_c$, in the unit of T)~\cite{ParaLimit}. As a demonstration, we show the $B_p$ value of sample W5
in Fig. 4(a), see the red dashed line. For each sample, the value of $B_{c2}^{\parallel}$ is always higher than $B_{c2}^{\perp}$, revealing the clear anisotropic effect in the thin films. In Fig. 4(b), we plot the
anisotropic parameter $\Gamma$ (=$B_{c2}^{\parallel}/B_{c2}^{\perp}$) as a function of the reduced temperature $t=T/T_c$. With the decrease of temperature, $\Gamma$ display a monotonous downward trend. This is
similar to that observed in Fe-based superconductors (Fe-based SCs)~\cite{Wang-12442-2020}.

To comprehensively evaluate the influences of the orbital pair-breaking effect, spin-paramagnetic pair-breaking effect, and spin-orbit scattering on the upper critical field, we checked the $B_{c2}$-$T$ data
based on the extended WHH theory~\cite{WHH-2}. According to this theory, the upper critical field in the dirty limit can be obtained from the function~\cite{WHH-2}
\begin{equation}
\begin{split}
ln\frac{1}{t}=&\sum_{\nu=-\infty}^{\infty}\{\frac{1}{|2\nu+1|}-[|2\nu+1|\\&+\frac{\bar{h}}{t}+\frac{(\alpha\bar{h}/t)^2}{|2\nu+1|+(\bar{h}+\lambda_{so})/t}]^{-1}\}, \label{eq:1}
\end{split}
\end{equation}
where $\bar{h}=4B_{c2}(T)/(\pi^2B^{\ast}T_c)$, $B^{\ast}=-dB_{c2}(T)/dT$$\mid$$_{T_c}$, and $\alpha$ and $\lambda_{so}$ are parameters reflecting the strength of the spin paramagnetic and spin-orbit scattering, respectively.
The parameter $\alpha$ is known as Maki parameter~\cite{Maki1964}. We first show the fitting results of sample W5 as an example. As shown in Fig. 5(a),
for the orientation of $B\perp$ film, the conventional scenario without spin related pair-breaking effects ($\alpha$ = 0) could not describe the experimental data (see the orange dashed line).
By adjusting the values of $\alpha$, the data can be represented by the theoretical model with the parameters $\alpha$ = 1.1, as shown by violet dashed line in Fig. 5(a).
In this fitting process, the spin-orbit scattering term $\lambda_{so}$ is fixed to zero. This means that the extended WHH model can describe the experimental data without considering the influence of
the spin-orbit scattering in the present system. The data of $B_{c2}^{\perp}$ for all the eight samples is shown in Fig. 5(c). It can be seen that the evolution of $B_{c2}^{\perp}$ with temperature for the eight samples can be
described by the extended WHH theory. The obtained Maki parameter $\alpha$ is summarized in Fig. 6 and will be discussed in the next section.

In the case of $B\parallel$ film, as shown in Fig. 5(b), the $B_{c2}^{\parallel}$-$T$ data reveals a very steep slope near $T_c$.
Similar behavior has also been observed in ultrathin SC materials~\cite{Ising,Ising-2,Ising-3},
which is attributed to the two-dimensional (2D) feature. According to the phenomenological Ginzburg-Landau (GL) model~\cite{GL}, in a 2D system, the upper critical field in the parallel direction should obey the equation
\begin{equation}
B_{c2}^{\parallel}(T)= B_{c2}^{\parallel}(0)\times(1-T/T_c)^{0.5}. \label{eq:2}
\end{equation}
The fitting result is represented by the red dashed line in Fig. 5(b). It can be seen that the 2D GL model can describe the experimental data in the low field region below about 6 T.
The deviation between the fitting curve and experimental data in the higher field region is a consequence of the changes in the dimensionality from 2D to 3D due to the decrease of the penetration depth with cooling.

\section*{4 Discussion}

\begin{figure}
\includegraphics[width=7cm]{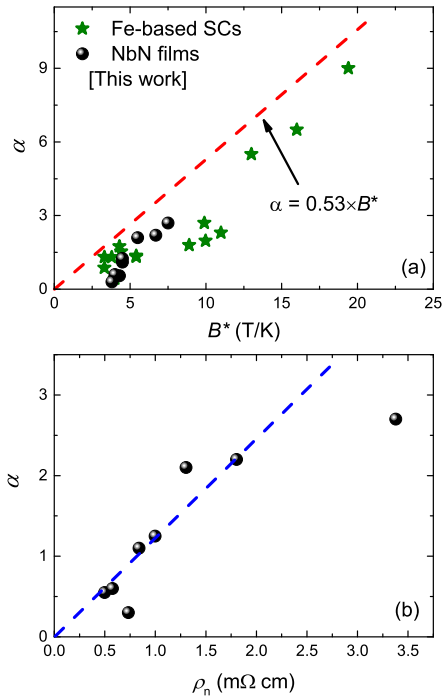}
\caption {(a) The Maki parameter $\alpha$ as a function of the slope $B^{\ast}$. The results from the Fe-based SCs are also shown for a
comparison~\cite{Yuan2009,Kano2009,Terashima2009,Lee2009,Fuchs2008,Khim2011,Fang2010,Khim2010,Xing2017,ZSWang2017,Wang-12442-2020}. The red dashed line denotes the Maki relation in the weak-coupling limit.
(b)The valued of $\alpha$ as a function of normal-state resistivity $\rho_n$. The blue line is a guide for eyes.} \label{fig6}
\end{figure}

Next, we checked the evolution of Maki parameter $\alpha$ with other parameters.
According the Maki formula~\cite{Maki1964}, $\alpha$ can be expressed as
\begin{equation}
\alpha=\frac{\sqrt{2}B_{c2}^{Orb}(0)}{B_{P}}, \label{eq:3}
\end{equation}
where $B_{c2}^{Orb}(0)=0.693\times B^{\ast}\times T_c$ is the upper critical field including only the orbital pair-breaking effect~\cite{WHH-1,WHH-2}. Considering the fact that $B_{P}=1.84\times T_c$,
a simple relation $\alpha=0.53B^{\ast}$ can be obtained~\cite{Wang-12442-2020}. This relation, which is represented by the red dashed line in Fig. 6(a),
is compared with the data obtained from the fitting using the extended WHH theory (Eq. 1). The results from the
Fe-based SCs~\cite{Yuan2009,Kano2009,Terashima2009,Lee2009,Fuchs2008,Khim2011,Fang2010,Khim2010,Xing2017,ZSWang2017,Wang-12442-2020} are also shown to have a comparison.
It is clear that, similar to the behavior of Fe-based system, the data of NbN films
also roughly follows the
tendency of the Maki formula. The departure of the data from the red dashed line has been attributed to the enhancement of $B_{P}$ due to the strong-coupling effect.
The similarity between the present NbN films and Fe-based SCs indicates that the two systems share a similar strength of SC coupling.
An important difference between the two systems is the field orientation: the Pauli paramagnetic effect and the related Maki parameter are observed in the in-plane field direction in Fe-based SCs, while this effect
emerges in the perpendicular direction in the NbN films.

In the dirty limit, the value of $\alpha$ is proportional to the $\rho_n$ if the normal state electronic specific heat coefficient $\gamma_n$ has no much change~\cite{WHH-2}.
In Fig. 6(b), we show the data of $\alpha$ as a function of $\rho_n$. Roughly speaking, $\alpha$ reveals a linear
dependence with $\rho_n$ as exhibited by the blue dashed line. Sample W11 depart clearly from this linear tendency. This behavior may suggest that the values of $\gamma_n$ is not a constant among these samples.
It is notable that the value of $\rho_n$ covers a wide range (0.5-3.4 m$\Omega$ cm), which is clearly larger than that of the single crystals of most sub-systems of Fe-base SCs~\cite{Wang-12442-2020,1111,122-K,122-Co}.
Considering the close correlation between $\rho_n$ and $\alpha$, we argue that the Pauli paramagnetic effect in the present NbN system originates from the enhanced normal-state resistivity.

The dimensional effect in the investigated thin films is another issue worthy of discussion.
In principal, dimensional effect can emerges with the decrease of film thickness~\cite{GL}. When the thickness is smaller than the penetration depth $\lambda$, the film can be penetrated by the parallel field, leading to reduction
of the diamagnetic energy as compared with the bulk superconductors.
In a film much thinner than the out-of-film coherence length $\xi_{\perp}$, the SC order parameter will not vary appreciably over the film.
An estimation from the upper critical field shows that the value of $\xi_{\perp}$ is no more than 10 nm in a wide temperature region, see Fig. S7 in SM.
The previous investigations have found that $\lambda$ is on the order of several hundred nanometers~\cite{Penetration1994,KAWAKAMI2004}. Thus, the thickness $d$ of the NbN films in this study (20-30 nm) is in the range $\xi_{\perp}<d<\lambda$.
As can be seen in Fig. 5(b), the $B_{c2}^{\parallel}$-$T$ data can be described by the 2D GL model in the temperature range
above 4.3 K. Moreover, the anisotropic parameter $\Gamma$ is below 3 for W5, see Fig. 4(b). This is also different from the conventional 2D systems with a large anisotropy.
The superconducting performances in the thin films with the thickness in this intermediate range is a subject worth studying in the future.

\section*{5 Conclusions}
In summary, we studied the upper critical field of high-quality NbN thin films, who's critical temperature and normal-state resistivity were tuned systematically by controlling the growth conditions.
The temperature dependence of $B_{c2}^{\perp}$ can be well described by the extended WHH theory, revealing the presence of Pauli paramagnetic effect in this system. The $B_{c2}^{\parallel}$-$T$ data near $T_c$
is consistent with the 2D GL model. The Maki parameter $\alpha$ shows a clear correlation with $-dB_{c2}(T)/dT$$\mid$$_{T_c}$ and $\rho_n$.
The large magnitude of $\rho_n$ may explain the observation of Pauli paramagnetic effect in the present system.

\section*{Data availability statement}
The data that support the findings of this study are available
upon reasonable request from the authors.

\section*{Acknowledgments}
This work is
supported by the National
Natural Science Foundation of China (No. 92065116), the Shanghai Technology Innovation Action Plan Integrated Circuit Technology Support Program (No. 22DZ1100200), and
the Key-Area Research and Development Program of Guangdong Province, China (No. 2020B0303030002). The authors would like to thank all the staff at the
Superconducting Electronics Facility (SELF) for their
assistance.

\section*{Supplementary Material}
The supplementary material is available online at http://xxx.






\end{document}